\documentclass[aps,pra,twocolumn,showpacs,superscriptaddress]{revtex4-1}
\usepackage{bbm}
\usepackage{graphicx}
\usepackage{amsmath}
\usepackage{amssymb,amsthm}
\usepackage{hyperref}
\usepackage[normalem]{ulem}
\usepackage{color}

\usepackage{mathtools}

\DeclarePairedDelimiter\floor{\lfloor}{\rfloor}

\newcommand{\ket}[1]{| #1 \rangle}

\newcommand{\mval}[1]{\langle #1 \rangle}
\newcommand{\eq}{\begin{eqnarray}} 
\newcommand{\en}{\end{eqnarray}}

\newcommand{\revisar}[1]{{\color{cyan}}}

\begin{document}

\title{Fermionic versus bosonic behavior of confined Wigner molecules}

\author{Eloisa Cuestas}
\affiliation{Universidad Nacional de C\'ordoba, Facultad de Matem\'atica, Astronom\'ia, F\'isica y Computaci\'on, Av. Medina Allende s/n, Ciudad Universitaria, X5000HUA C\'ordoba, Argentina}
\affiliation{Instituto de F\'isica Enrique Gaviola (IFEG), Consejo de Investigaciones Cient\'ificas y T\'ecnicas de la Rep\'ublica Argentina (CONICET), C\'ordoba, Argentina}

\author{P. Alexander Bouvrie}
\affiliation{Departamento de Estadística e investigación operativa, Universidad de Granada, 18071-Granada, Spain}

\author{Ana P. Majtey}
\affiliation{Universidad Nacional de C\'ordoba, Facultad de Matem\'atica, Astronom\'ia, F\'isica y Computaci\'on, Av. Medina Allende s/n, Ciudad Universitaria, X5000HUA C\'ordoba, Argentina}
\affiliation{Instituto de F\'isica Enrique Gaviola (IFEG), Consejo de Investigaciones Cient\'ificas y T\'ecnicas de la Rep\'ublica Argentina (CONICET), C\'ordoba, Argentina}

\date{\today}


\begin{abstract}
We assess whether a confined Wigner molecule constituted by $2N$ fermions behaves as $N$ bosons or $2N$ fermions.   
Following the work by C. K. Law [Phys. Rev. A \textbf{71}, 034306 (2005)] and Chudzicki et al. [Phys. Rev. Lett. \textbf{104}, 070402 (2010)] we discuss the physical meaning and the reason why a large amount of entanglement is needed in order to ensure a bosonic composite behavior. By applying a composite boson ansatz, we found that a Wigner molecule confined in two dimensional traps presents a bosonic behavior induced by symmetry. The two-particle Wigner molecule ground state required  by the composite boson ansatz was obtained within the harmonic approximation in the strong interacting regime. Our approach allows us to address few-particle states (widely studied within a variety of theoretical and numerical techniques) as well as a large number of particles (difficult to address due to computational costs). For a large number of particles, we found strong fermionic correlations exposed by the suppression of particle fluctuations. For a small number of particles, we show that the wave function calculated within the composite boson ansatz captures the Friedel-Wigner transition. The latter is shown in a regime in which strong correlations due to the Pauli exclusion principle arise, therefore, we conclude that the coboson ansatz reproduces the many particle physics of a confined Wigner molecule, even in the presence of strong deviations of the ideal bosonic behavior due to fermionic correlations. 

\end{abstract}

\maketitle

\section{Introduction}
\label{sec_intro}

In our Condensed Matter Physics, Solid State Physics or Modern Physics courses, we learn to deal with composite particles (such as atoms, molecules, and nuclei) made up of an even number of fermions as ideal bosons. This strategy is widely used in Many Particle Physics and appears to be justified due to the success of such approach. Just to mention a few examples, it has allowed us to describe atomic Bose-Einstein condensates, excitons, and Cooper pairs in superconductors \cite{Chudzicki2010, TichyBouvrie2014,Keldysh1968,Shumway2001,Golomedov2017}. In this context, it is natural to ask under which conditions a pair of fermions can be treated as an elementary boson. Many authors have explored this question, in particular, C.K. Law pointed out that the key to understand the origin of the composite behavior relies on entanglement: the composite particles can be treated as bosons as long as they are sufficiently entangled. Law’ s hypothesis also expresses that mechanical binding forces are not essential when applying the composite representation, in other words, such forces would act only to enforce quantum correlations \cite{Law2005}. Here we present new evidence in this direction by characterizing the bosonic degree of the finite-size analogue of Wigner crystals, namely Wigner molecules \footnote{When considering a system of $M$ confined particles interacting via some repulsive potential, the tradeoff or competition between repulsion and confinement may yield to a regime in which the particles arrange into $M$ positions, giving rise to a correlated state called \textit{Wigner molecule}. It is well known that when Coulomb interactions are sufficiently strong, the ground state of a one- or two-dimensional quantum dot is a Wigner molecule.}. Within the framework of the composite bosons ansatz \cite{Law2005, Leggett2001,CombescotTanguy2001}, we calculate the density of states, occupations probabilities, and the particle density profile of a Wigner molecule made up of an even number of fermions. We show that a Wigner molecule confined in two dimensional traps presents an ideal bosonic behavior induced by symmetry. We support Law’s hypothesis by showing that even for large interaction strengths, the system presents deviations from purely bosonic behavior when the amount of entanglement is not sufficiently large. We also discuss the connection exposed by Chudzicki et al. in Ref. \cite{Chudzicki2010} between the space available to each composite particle and the condition given by Law in Ref. \cite{Law2005} to guarantee the validity of the bosonic particle description, i. e. the effective number of Schmidt modes must be much greater than the total number of composite particles.  

We focus on Wigner molecules (WMs) as a realistic quantum system depicting strong electronic correlations, which have been experimentally observed in a wide variety of systems, such as two-dimensional semiconductor heteroestructures \cite{andrei_1998,piot_2008}, semiconductor quantum dots \cite{kallikaos_2008}, 
one-dimensional quantum wires \cite{ellenberger_2006,singha_2010,kristinsdottir_2011,meyer_2009}, carbon 
nanotubes \cite{deshpande_2008,pecker_2013}, and in crystalline states for dusty plasma \cite{melzer_2003}. In order to describe and understand these molecules, several theoretical approaches have been implemented \cite{Cavaliere2014}. The theoretical studies can be divided into two groups: analytical methods (such as the Bethe ansatz \cite{Xianlong2012, Wang2012}, Luttinger liquids \cite{Mantelli_2012, Gambetta_2014}, or solving the Schr\"odinger equation \cite{koscik_2015, Cuestas2017, Koscik2017}) and methods involving strong numerical calculations (such as density functional theory \cite{Xianlong2012, Wang2012}, lattice regularized diffusion quantum Montecarlo methods \cite{Casula2006}, path integral Monte Carlo methods \cite{Kylanpaa2016}, and exact diagonalization \cite{Secchi2010, Secchi2012}). The experimental proof of the existence of Wigner molecules relies on the observation of a sharp collapse of the energy gap between the ground and the first excited state, which can be detected by performing transport measurements \cite{Cavaliere2014}. In Ref. \cite{pecker_2013}, by using transport spectroscopy of ultraclean nanotube quantum dots the authors provide an unambiguous demonstration of the presence of the strongly interacting quantum ground state of a Wigner molecule. The results of those experiments were analyzed by comparing them with exact-diagonalization calculations, which allowed to identify changes in the symmetry of the state as a fundamental signature of a strongly interacting quantum Wigner molecule.

In the present work we propose a new approach to study a WM made up of an even number of constituents: to apply the composite boson ansatz to $N$ pairs of confined particles. The underlying idea is to include exchange correlations in the widely (and routinely) used strategy of treating composite particles made of an even number of constituents as non-interacting bosons. The basic unit used to describe the system is a composite particle made up of two particles of two distinct and distinguishable species. Since the particles of the same species are indistinguishable, the ansatz takes into account symmetry requirements as well as the Pauli exclusion principle when it is applied to fermionic systems. When taking the coboson as a unit, a system made up of $N$ of such entities will necessarily have $2N$ particles. Therefore, addressing a $2N$ unpolarized system in the framework of the coboson ansatz is natural while addressing a polarized system or a system made up of an odd number of particles is, at least, a non-straightforward issue (it would demand an extension of the ansatz -which to the best of our knowledge has never been addressed- or the use of other techniques requiring strong numerical calculations -such as density functional theory or quantum Montecarlo methods \cite{Xianlong2012, Wang2012, Kylanpaa2016}-). The coboson ansatz is valid for any number of pairs of particles and leads to a considerable improvement of the computational time of the many-particle state and observables of interest when comparing to the computational time required by other techniques such as Quantum Monte Carlo methods. It allows to address few particles states (widely studied, see for example \cite{Xianlong2012, Wang2012, Soffing2012, Secchi2012, Kylanpaa2016}) as well as many particle states, or even large ensembles of particles (see for instance Ref. \cite{Bouvrie2019} were we study $10^3$ fully entangled fermionic atoms). We present our results for a small number of pairs, $10^2$ or less, far more than the number of particles treated in previous works, \cite{Xianlong2012, Wang2012, Soffing2012, Secchi2012, Kylanpaa2016}. We hope this facts will enhance the value of our study, especially when taking into account that the computational complexity involved in the study of Wigner molecules scales exponentially with $N$ (actually, the numerical diagonalization method becomes very inefficient already for $5$ pairs of fermions, even in dimension one \cite{Kylanpaa2016}). Here, by means of the coboson ansatz we are able to deal with a large number of pairs with low computational cost. The latter advantage allowed us to observe the presence of strong fermionic correlations exposed by the suppression of particle fluctuations, as was recently reported for ultracold Fermi gases \cite{Bouvrie2019}. Finally, we also show that the coboson ansatz is capable to capture the Friedel-Wigner crossover described in Refs. \cite{Xianlong2012, Kylanpaa2016}.

This article is organized as follows. We review our framework, the composite boson ansatz, in section~\ref{sec_ansatz}. In section~\ref{sec_WM} we discuss the physical reason why a large degree of entanglement between the constituents of the composite particle ensures that the two particles can be treated as elementary bosons and give more evidence supporting Law's hypothesis. Sections~\ref{sec_res_1D} and \ref{sec_res_2D} attempt to show that the coboson ansatz can be a powerful and useful approach in the study of Wigner molecules or similar systems. Section~\ref{sec_res_1D} is devoted to the analysis of the bosonic degree of a WM confined in one dimension by studying the density of states, occupations probabilities, and the ground state particle density profile. The proof of the ideal bosonic behavior induced by symmetry in a system of $N$ pairs of fermions confined in two (and higher) dimensions is presented in section~\ref{sec_res_2D}. Finally, a summary is given in section~\ref{sec_concl}.

\section{The Coboson ansatz}
\label{sec_ansatz}

In order to address a Wigner molecule made up of an even number of fermions we draw upon the composite boson (or coboson) ansatz. This formalism was introduced in 2001 to study correlated pairs of particles. A. J. Legget developed the ansatz for bosons \cite{Leggett2001}, and M. Combescot and C. Tanguy  presented it for correlated pairs of fermionic particles \cite{CombescotTanguy2001}. The formalism was extensively applied to describe fermionic systems, such as excitons which feature long range Coulomb interactions \cite{CombescotShiau2015}, ultracold interacting Fermi gases \cite{CombescotShiauChang2016, BouvrieTichyRoditi2016, Bouvrie2019}, and Cooper pairs \cite{Pong2007}. When applied to the Bose-Einstein Condensate regime of ultracold interacting Fermi gases this formalism gives an universal dimer-dimer scattering length $a_{dd}^\text{Cob}=0.64\, a$ \cite{ShiauCombescotChang2016}, in great agreement with the exact solution $a_{dd}\approx0.6\, a$ \cite{PetrovShlyapnikov2004}. Another achievement obtained within this framework concerns the calculation of the molecular condensate fraction \cite{BouvrieTichyRoditi2016}, which matches the Bogoliubov results and fixed-node diffusion Monte Carlo \cite{Giorgini2005}. 

In what follows we summarize the key point of the formalism as presented by C.K. Law in Ref. \cite{Law2005}. We consider a composite particle formed by two distinguishable particles (both bosons or fermions) of distinct species $a$ and $b$. By solving a suitable two-particle Hamiltonian we are able to obtain the two body ground state wave function $\Psi\left( x_a, x_b\right)$. We can express this wave function in its Schmidt decomposition form,

\eq
\label{eq_schmidt_form}
\Psi \left( x_a, x_b \right) = \sum_j^{\infty} \sqrt{\lambda_{j}} \phi_{j}^{(a)}(x_a) \phi_{j}^{(b)}(x_b) ,
\en 

\noindent where $\phi_{j}^{(a)}$ ($\phi_{j}^{(b)}$) are the natural orbitals (Schmidt modes) and $\lambda_{j}$ are their occupations (Schmidt numbers) \cite{fedorov_2014}. These orbitals give a complete and orthonormal set for the Hilbert space of each particle and are the eigenvectors of the one-particle reduced density matrix \cite{Law2005, Cuestas2017}, with the occupations as the associated eigenvalues. As it is well-known, the entropies usually applied when quantifying entanglement are calculated in terms of this occupancies \cite{Law2005, Cuestas2017}.      

The coboson ansatz states that in second quantization the $N$-coboson pair-correlated state $\ket{N}$ is given by successive applications of identical coboson creation operators $\hat c^\dagger$ acting on the vacuum $\ket{0}$ \cite{CombescotMatibetEtal2008},

\eq
\label{eq_ket_N}
\ket{N} =\frac{1}{\sqrt{N! \chi_{N}}} \left(\hat c^\dagger \right)^N \ket{0} .
\en 

\noindent The operator $c^\dagger$ creates two particles in the entangled state defined in Eq.~\ref{eq_schmidt_form}, hence we can write this operator as 

\eq
\label{eq_c_dagg}
c^\dagger = \sum_j^{\infty} \sqrt{\lambda_{j}} a_j^\dagger b_j^\dagger ,
\en 

\noindent where $a_j^\dagger$ ($b_j^\dagger$) is the creation operator of a particle of the species $a$ ($b$) in the Schmidt mode or natural orbital $\phi_{j}^{(a)}$ ($\phi_{j}^{(b)}$). The $\chi_{N}$ factor is a normalization constant introduced in order to obtain $\langle N \vert N \rangle = 1$. As stated by Law in Ref. \cite{Law2005}, the two distinguishable particles act as an ideal composite boson when the ratio $\chi_{N+1}/\chi_N \to 1$ (under this condition the operators $c^\dagger$ and $c$ obey the bosonic commutation relation and the system of $N$ pairs of particles of distinct species can be regarded as $N$ independent non-interacting bosons). In other words, the ratio of these normalization constants reveals the deviations of the ideal bosonic behavior. The normalization constant for two distinguishable bosons ($\chi_N^B$) or fermions ($\chi_N^F$) can be written as

\eq
\label{eq_chi_B}
\chi_{N}^B = N! \sum_{j_1 \leqslant j_2 \leqslant \ldots \leqslant j_N} \lambda_{j_1} \lambda_{j_2} \ldots \lambda_{j_N} ,
\en 

\eq
\label{eq_chi_F}
\chi_{N}^F = N! \sum_{j_1 < j_2 < \ldots < j_N} \lambda_{j_1} \lambda_{j_2} \ldots \lambda_{j_N} .
\en 

\noindent Our main focus will be on the fermionic case, nevertheless, we have also performed some calculations for bosonic particles and thus we give a general guideline for the coboson ansatz formulation. 

From Eqs.~\eqref{eq_chi_B} and \eqref{eq_chi_F} we can see that the normalization factors contains the information about how the particles distribute over the available single-particle states or natural orbitals associated to the two-particle state of Eq. (\ref{eq_schmidt_form}) \cite{BouvrieTichyMolmer2016}. On the one hand this means that our description of the $2N$-particle system relies heavily on the behavior of (and availability of close approximations to) the natural orbitals. On the other hand  this description reveals the quantum correlations by showing explicitly the pairing structure (from Eq.~\eqref{eq_schmidt_form} we can notice that when the particle $a$ is in the mode $\phi_{j}^{(a)}$, then the particle $b$ must be in the mode $\phi_{j}^{(b)}$) and it allows to obtain observables in terms of the single-particle states, turning the results more intuitive and accessible.

In the fermionic case the particle distribution among the single-particle states follows the Pauli exclusion principle, in this sense the ansatz only considers the interaction between fermions given by fermionic exchanges \cite{CombescotMatibet2010}. Then, the physics of the many-particle system comes up from the state of a single pair given by Eq.~\eqref{eq_schmidt_form} and from the fermionic exchange interactions, both of them regarded in $\chi_N^F$. From Eq.~\eqref{eq_ket_N} and \eqref{eq_c_dagg}, together with the Pauli exclusion principle (i.e. $( \hat a_{j}^\dagger)^2 = (\hat b_{j}^\dagger)^2=0$), we can write the state for $N$ cobosons made up of pairs of fermions as

\eq
\label{eq_ket_N_F}
\ket{N} = \frac{1}{\sqrt{N!\chi_N^F}} \sum_{\left\lbrace j_1,j_2, \ldots, j_N \right\rbrace^{'}} \left(\prod_{k=1}^N \sqrt{\lambda_{j_k}} \hat a_{j_k}^\dagger \hat b_{j_k}^\dagger \right) \ket{0},
\en 

\noindent with $\left\lbrace j_1,j_2, \ldots, j_N \right\rbrace^{'}$ indicating that the sum is over all the indices appearing in the summand, subject to the restriction that the indices $j_1, j_2,\ldots,j_N$ must take distinct values. This guarantees the Pauli exclusion principle: there can not be two identical fermionic particles in the same quantum state. 

As we have already mentioned, this representation of the state plays a crucial role in the improvement of the computational cost. The calculation of the coboson state and related observables require short computational times \cite{BouvrieTichyRoditi2016,Bouvrie2019}, and the analytical expressions used and developed here are relatively simple (in the sense that they do not require further quantum field knowledge). 

The calculation of the normalization factor becomes crucial not only because it is involved in the definition of the state $\ket{N}$ but because it (or a very similar form) appears in the expressions obtained for many observables of interest \cite{Bouvrie2019}. In practice closed calculations of $\chi_N$ are rare and one must to befriend with the elementary symmetric polynomial theory (see section 2 of Ref. \cite{TichyBouvrie2014} about the connection between $\chi_N$ and these polynomials) in order to exploit some recursion identities,

\eq
\label{eq_rec_chis}
\chi_{N}^F = \sum_{m=1}^{N} \frac{(N-1)!}{(N-m)!} (-1)^{m+1} \chi_{N-m}^F M(m),
\en 

\noindent with $M(m) = \sum_{n} \lambda_n^m$ being the power sum of order $m$ (notice that these power sums can be related to the R\'enyi entropy of the distribution of $\lambda_n$ \cite{TichyBouvrie2012a}) and setting $\chi_{0}^F = 1$ for convenience. A non-recursive expression \cite{boklan2018}, not used (to our knowledge) in the context of the coboson ansatz so far, states

\begin{eqnarray}
\label{eq_chis_sums}
\chi_{N}^F = N! \sum_{\left\lbrace k_1, k_2, \ldots, k_N \right\rbrace^{''}} &&\frac{(-1)^{k_1 + k_2 + \ldots +k_N + N}}{k_1!k_2! \ldots k_N!} \\
\nonumber
&&  \prod_{i=1}^{N} \left( \frac{M(i)}{i} \right)^{k_i} ,
\end{eqnarray} 

\noindent where $\left\lbrace k_1, k_2, \ldots, k_N \right\rbrace^{''}$ indicates that the sum is over all the indices appearing in the summand, with the restriction $k_1 + 2 k_2 + \ldots + N k_N = N$. This expression will be of particular interest when addressing two-dimensional Wigner molecules.    


\section{Entanglement's key role in ensuring the ideal bosonic behavior of composite particles} 
\label{sec_WM}

We begin by computing the occupations and entanglement of a two-particle Wigner molecule by solving the Schr\"odinger equation of two distinguishable interacting particles confined in a two-dimensional anisotropic harmonic trap (see appendix~\ref{sec_appendix_2pWM}). By varying the anisotropy parameter one can obtain the results for the two dimensional isotropic case as well as the one dimensional situation. Since Wigner localization is expected for low density systems or for large interaction strengths, we focus in the strong interacting regime. The procedure is quite simple: when introducing the center of mass and relative coordinates the two-particle Hamiltonian decouples and the total ground state wavefunction calculated within the harmonic approximation in the strong interacting regime is the product between two harmonic oscillators ground states \cite{Cuestas2017}. Once we have the total ground state we are able to write it in the form of Eq.~\eqref{eq_schmidt_form}, with explicit reference to the natural orbitals and its occupations. 

The occupation numbers for the two dimensional case are products of the form 

\eq
\label{eq_occup}
\lambda_{j_x, j_y}^{2D} = \lambda_{j_x} \lambda_{j_y} = (1-z_x) z_x^{j_x} (1-z_y) z_y^{j_y} ,
\en 

\noindent where in the interest of coherence we maintain the notation introduced by Law in Ref. \cite{Law2005}. Both parameters $z_x$ and $z_y$ are defined in the range $0<z_{x,y}<1$, and $j_{x,y} \geqslant 0$ are integers. As we show in appendix~\ref{sec_appendix_2pWM}, $z_x$ depends on the form and parameters of the interaction potential as well as on the ratio between the interaction strength and the confining energy, while $z_y$ depends on the anisotropy of the trap. Notice that the one dimensional case is re-obtained by taking $z_y \to 0$ and the isotropic trap is obtained when $z_y \to 1$(detailed discussion, calculations and expressions are given in appendix~\ref{sec_appendix_2pWM}). 

In Ref. \cite{Cuestas2017}, it is shown that due to the product form of the occupancies, the von Neumann, min-entropy, max-entropy, and R\'enyi entropies are a sum of terms associated to each one of the two factors appearing in the two-dimensional occupation numbers Eq.~\eqref{eq_occup}, where one of these terms depends on the anisotropy parameter and the other term is associated to the interaction potential. Having the explicit expressions for these entropies (see appendix~\ref{sec_appendix_2pWM}) it is easy to see that in the one dimensional case the von Neumann, min-entropy, and the family of R\'enyi entropies diverge if $z_x \to 1$, case in which the linear entropy tends to one (its maximum value). These results support the idea stated by Law and discussed by Chudzicki et al. in Ref. \cite{Chudzicki2010} about the importance of entanglement when treating composite particles, made up of bosons or fermions, as bosons. In Ref. \cite{Law2005} the author calculates the Schmidt number $\cal{K}$ (inverse of the purity) as entanglement measure and concludes that the composite representation can be applied to strongly entangled particles, which are not limited to mechanically bounded systems. Here we establish the connection between his results and the divergence of a variety of entanglement entropies found in Ref. \cite{Cuestas2017}, in which the authors also show that large interaction strengths do not implicate a large degree of entanglement between the constituents of the WM. In other words, strongly mechanically bound particles present deviations from ideal composite boson character if their degree of entanglemet is poor, as stated by Law.

Law's conclusion is that the bosonic particle description is valid when the effective number of Schmidt modes is much greater than the total number of composite particles, i.e. when there is enough place in the space of states to accommodate all the pairs. As it is shown by Baccetti and Visser in Ref. \cite{Baccetti_2013}, large von Neumann entropies come from exponentially large regions of state space. Then, in order to obtain an infinite von Neumann entropy an infinite number of states must have non-zero probability. This means that the divergences found in Ref. \cite{Cuestas2017} in the von Neumann, min-entropy, and the family of R\'enyi entropies endorse the analysis made by Law. 

Now we would like to introduce a new twist in the discussion. In Ref. \cite{Cuestas2017} it is also shown that in two (or higher) dimensional case, no matter the range or strength of the interaction potential between the constituent parts of the molecules, the entropies remain finite when considering an anisotropic trap and diverge when varying the symmetry of the trap towards the isotropic case. These divergences in the entanglement entropies indicate that when reaching the isotropic trap the two distinguishable particles can be described as an elementary boson, therefore, we are in the presence of an ideal bosonic behavior induced by symmetry. We will return to this point in Sec.~\ref{sec_res_2D} by showing that the normalization ratio $\chi_{N+1}/\chi_N$ goes to one if $z_y \to 1$ no matter the form, range or strength of the interaction potential, i.e. for any value of $z_x$.

Before concluding, we would like to cover one last issue. By calculating the purity of a confined hydrogen atom, the authors of Ref. \cite{Chudzicki2010} establish the equivalence between Law's condition (which states that the effective number of Schmidt modes must be greater than the total number of composite particles in order to guarantee the suitability of the composite boson description) and the condition that the available space to each pair must be large compared to its size. Following the procedure of Ref. \cite{Cuestas2017} it is possible to calculate the ratio between the available space (given by the trap) and the space occupied by the pair (given by the parameters associated to the potential interaction via the harmonic approximation). Our results support the equivalence between Law's and Chudzicki's conditions: we found that the available space for each pair is large compared to its size only when all the calculated entropies are sufficiently large. At this point it is important to keep in mind that the ground states considered by Law, Chudzicki et al., and ourselves in this argument have the same form and correspond to confined particles. It is reasonable to expect this equivalence to be valid when treating confined particles because the pairs not only need sufficient space in the state space as pointed by Baccetti and Visser in Ref. \cite{Baccetti_2013}, but also enough available real space compared to its size. This logic leads immediately to the question if this is a general equivalence or if it only holds in the context of confined particles, question which remains open and requires further discussion.   

\section{One dimensional Wigner molecules} 
\label{sec_res_1D}

\begin{figure}[]
\includegraphics[width=0.7\columnwidth]{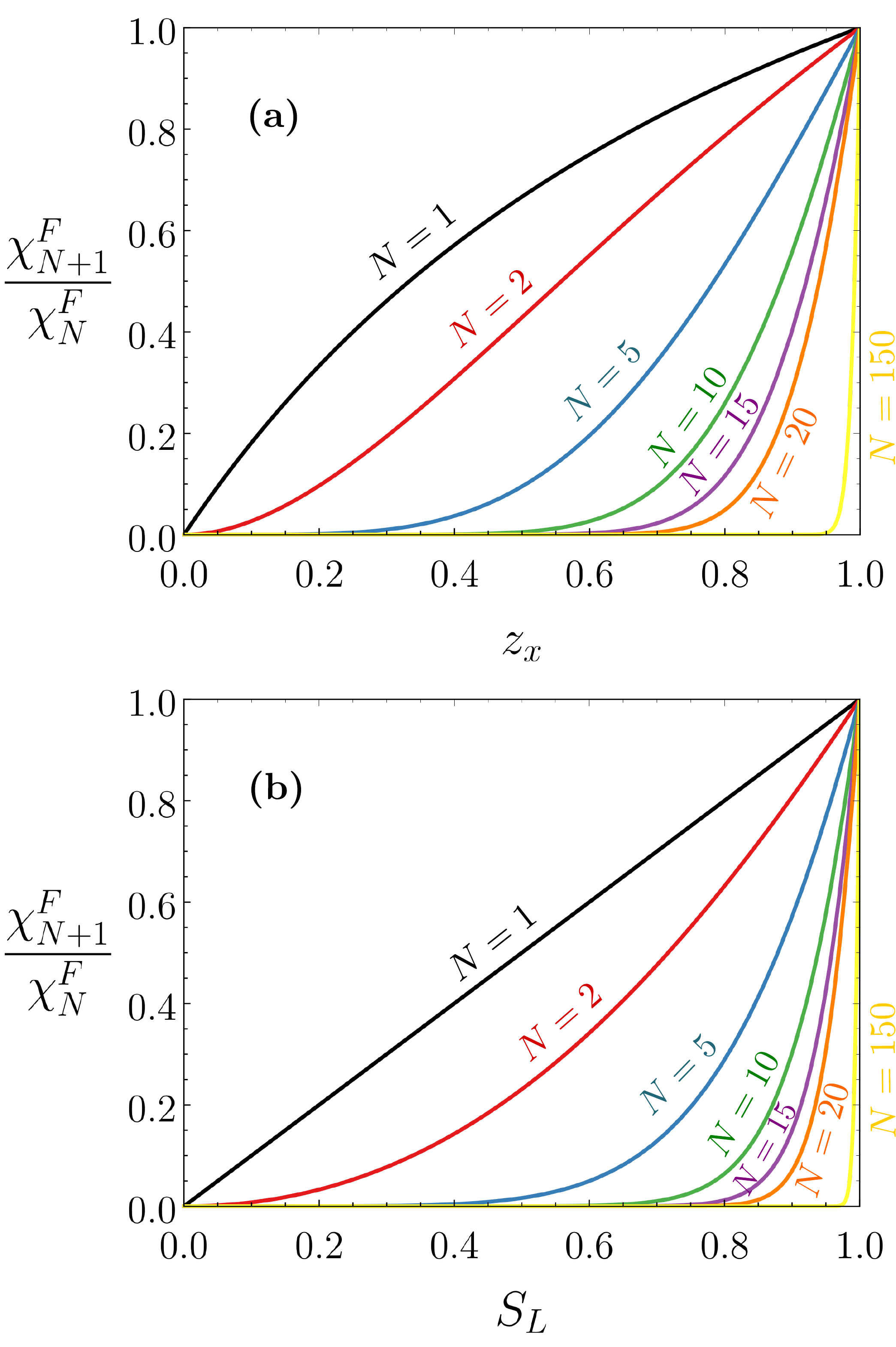}
\caption{The deviations of the ideal bosonic behavior are reduced when the degree of entanglement between the constituents of the fermionic pairs increases, here revealed by the normalization ratio $\chi_{N+1}^F/\chi_N^F$ plotted as a function of the parameter $z_x$ in (a) and as a function of the linear entropy $S_L$ in panel (b) for $N=1,\,2,\,5,\,10,\,15,\,20,\,150$ from top to bottom in black, red, blue, green, purple, orange, and yellow solid lines respectively.}
\label{fig_chi_ratio_F_1D}
\end{figure}

In the present section we characterize the bosonic degree of a one-dimensional Wigner molecule made up of an even number of fermions. For a few number of fermionic pairs, this system has been widely studied within a vast variety of methods \cite{Xianlong2012, Wang2012, Soffing2012, Secchi2012, Kylanpaa2016}. Here we show our results for few ($N\leqslant5$) and many ($N \thicksim 10^2$) pairs of fermions. Our approach can be summarized as follows: first, we need to solve the two-particle Hamiltonian to obtain the two-particle state of the pair of fermions that constitute a single composite boson (see appendix~\ref{sec_appendix_2pWM}), then, the total state of a WM made up of $2N$ fermions is given by the $N$-coboson correlated state $\ket{N}$, which is calculated as stated in Eq.~\eqref{eq_ket_N_F}. The strategy proposed by Law in Ref. \cite{Law2005} can be used to compute the normalization constant $\chi_N^F$, as well as other quantities of interest that can be defined in terms of other symmetric polynomials analogous to the normalization factor. As mentioned in Sec.~\ref{sec_WM} the linear, von Neumann, min-entropy, and the family of R\'enyi entropies diverge when $z_x \to 1$ (see also appendix~\ref{sec_appendix_2pWM}), case in which the normalization ratio $\chi_{N+1}^F/\chi_N^F$ goes to one (Fig.~\ref{fig_chi_ratio_F_1D}).  The normalization ratio as a function of the linear entropy $S_L$ is depicted in Fig.~\ref{fig_chi_ratio_F_1D} (b), showing that $\chi_{N+1}^F/\chi_N^F \to 1$ when $S_L \to 1$. Therefore, the deviations of the ideal bosonic behavior are reduced when the degree of entanglement between the constituents of the fermionic pairs increases. Fig.~\ref{fig_chi_ratio_F_1D} also shows that the ideal bosonic behavior is attenuated faster for an increasing number of particles: the normalization ratio decays more rapidly to zero for larger $N$.

\begin{figure}[]
\includegraphics[width=0.65\columnwidth]{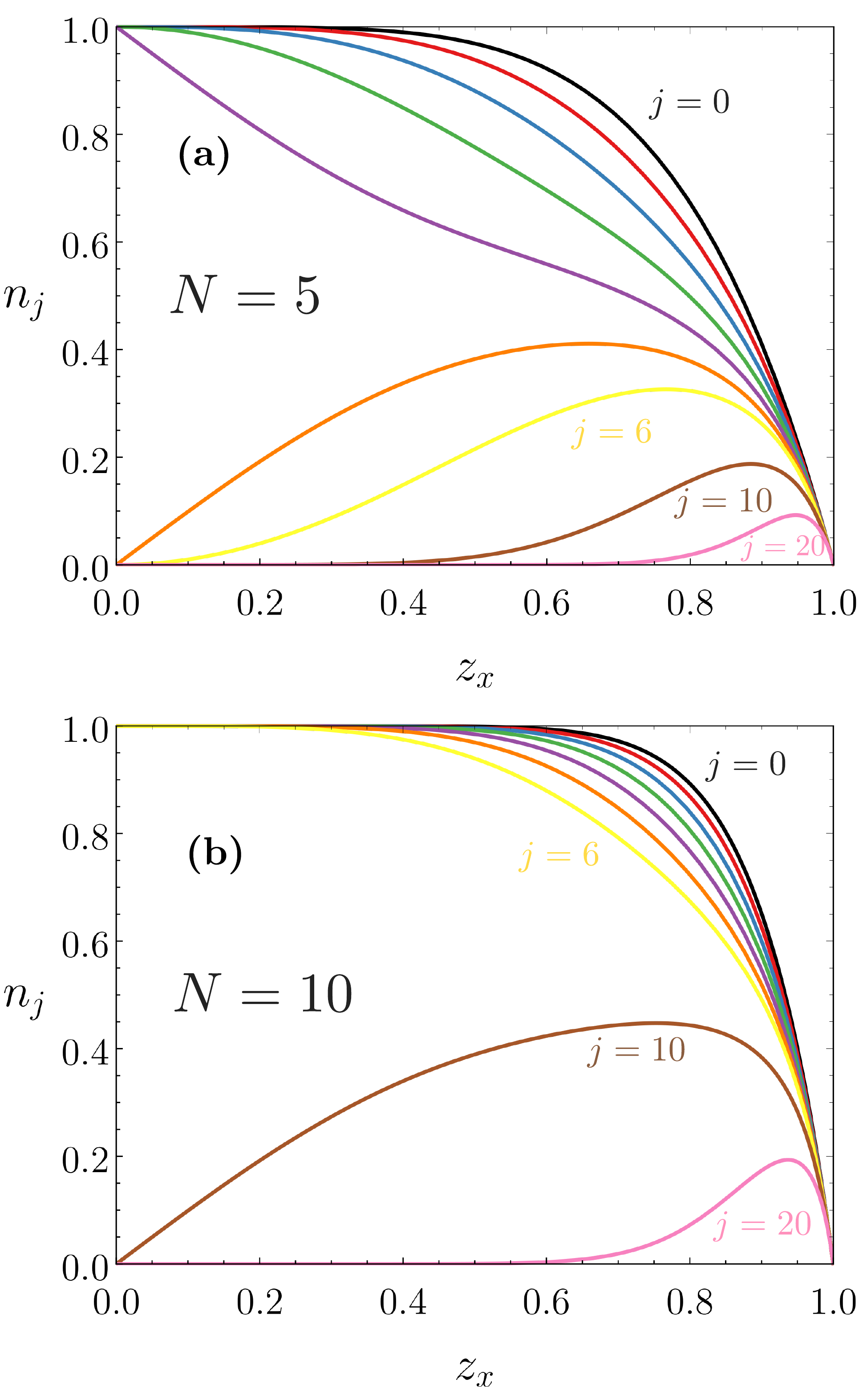}
\caption{Populations $n_j$ of the first seven single-particle states $\phi_{j}$ as functions of $z_x$ for $N=5$ (a) and $N=10$ (b) particle pairs. $j=0,\,1,\,2,\,3,\,4,\,5,\,6,\,10,\,20$ are shown from top to bottom in black, red, blue, green, purple, orange, yellow, maroon, and pink solid lines respectively.}
\label{fig_populations}
\end{figure}

The populations of the single-fermion states can be writen as $n_j(N)=N \lambda_j \chi_{N-1}^{\lambda_j}/\chi_N$, where $\chi_{N-1}^{\lambda_j}$ are the normalization factor calculated without considering the occupation $\lambda_j$ \cite{Bouvrie2019}, and the superscript $F$ of fermions has been omitted for simplicity of notation. They can be obtained in a closed form or evaluated by using the recursion formula given in Eq.~\eqref{eq_rec_chis}. The first $N$ states are fully occupied for a wide range of $z_x$ (i.e. $n_j \thicksim 1$ for $j=0,\,1,\,\ldots,\,N-1$), range which gets broader when increasing $N$ (see Fig.~\ref{fig_populations}). All the populations vanish for $z_x=1$ and the populations with $j \geqslant N$ vanish also for $z_x=0$, as can be seen in figure~\ref{fig_populations}. 

\begin{figure}[]
\includegraphics[height=0.77\textheight]{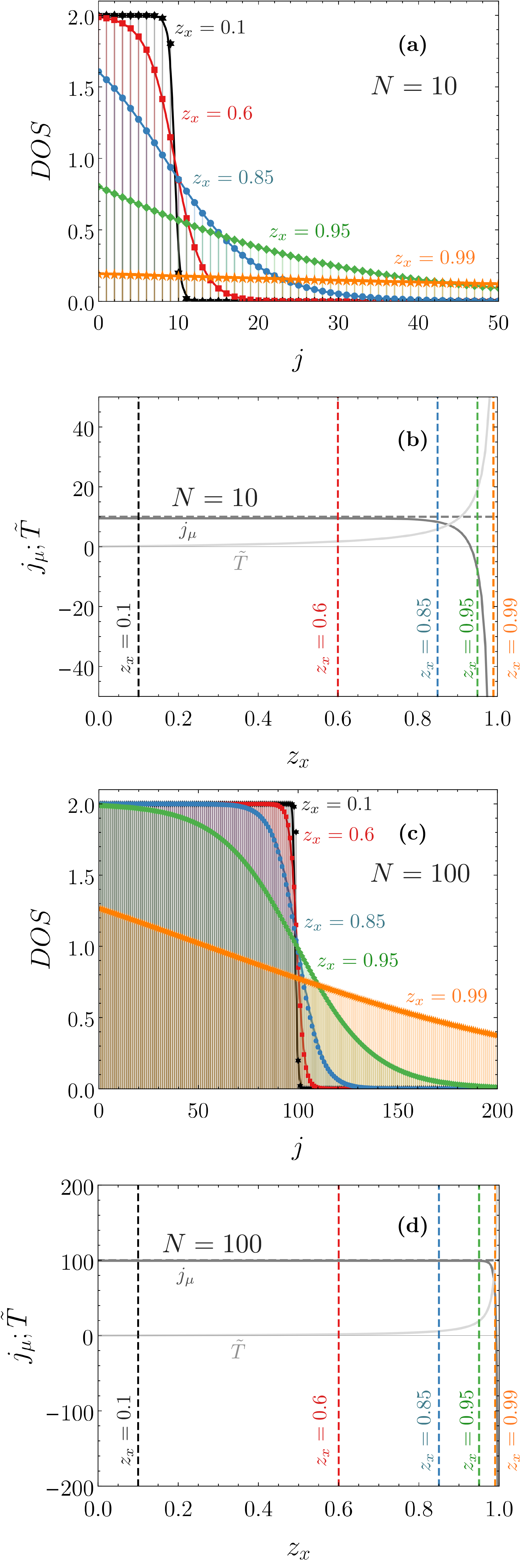}
\caption{Bosonic vs. fermionic character of the $N$ confined particle pairs exposed by the density of states (DOS) for $N=10$ (a) and $N=100$ (c). For small values of $z_x$ the obtained step-function reproduces the Fermi-Dirac distribution. For values of $z_x$ close to one all the single-particle states have a small non-zero population and the $N$ particle pairs of distinct species behave like $N$ ideal bosons. The values $z_x = 0.1,\, 0.6,\,0.85,\,0.95,\,0.99$ are depicted from top to bottom in panels (a) and (c) in black six-pointed stars, red squares, blue dots, green diamonds, and orange five-pointed stars respectively. The effective temperature $\tilde{T}$ and the value of $j$ associated to the Fermi energy $j_{\mu}$ (obtained by fitting the DOS with the Fermi-Dirac distribution) are depicted for $N=10$ in (b) and for $N=100$ in (d).}
\label{fig_DOS}
\end{figure}

Once we have the populations of each single-fermion state $\phi_{j}$, it is possible to calculate the density of states (DOS) of the $N$ confined particle pairs as the population multiplied by the degeneracy of the single-particle state, $DOS = g_j n_j$. In order to obtain the DOS we need to be able to associate each $\phi_{j}$ with its energy. By rewriting the total two-particle Hamiltonian derived after performing the harmonic approximation of the relative Hamiltonian \cite{Cuestas2017}, a sum of two one-particle Hamiltonians plus two crossed terms is obtained. The expectation values of the crossed terms vanish for states of the form $\phi_{j}^{(a)}(x_a) \phi_{j}^{(b)}(x_b)$ and the expectation values of the two one-particle Hamiltonians are the energies of a single-particle oscillator. Therefore, the energies of the single-particle states $\phi_{j}$ are proportional to $j$ and we can consider equivalently to increase the energy or increase $j$. Then, we are able to depict the DOS as a function of $j$, as shown in figure~\ref{fig_DOS}. When increasing the parameter $z_x$ the DOS changes smoothly from a step-function behavior (centered in $N$, the number of pairs of fermions) to twice the occupation number of the $j$-state, i.e. $2 (1-z_x) z_x^j$. This means that for small values of $z_x$ the $N$ lowest energetic states are occupied while all the remaining states's populations are null. For larger values of $z_x$ the particles occupy states with $j\geqslant N$ and for $z_x$ sufficiently close to unity all the states are occupied with non-zero populations, in agreement with our previous analysis of the populations $n_j$ (Fig.~\ref{fig_populations}). 

The behavior of the DOS reveals the presence of strong fermionic correlations. For $z_x$ sufficiently small, the obtained step-function reproduces the Fermi-Dirac distribution. In this sense, an increase in $z_x$ breaks the perfect step-function resembling what the temperature does to the Fermi distribution. For $z_x$ sufficiently close to unity all the single-particle states have a small but non-zero population, the entropies diverge as discussed in Sec.~\ref{sec_WM}, and the $N$ pairs of fermions of distinct species behave like $N$ ideal bosons.

By fitting the obtained density of states with the expression given by the Fermi-Dirac (FD) distribution, we are able to find an effective temperature of the fermionic system. We fitted the calculated DOS to the function $g_j / \left( e^{(j-j_{\mu})/\tilde{T}} +1 \right) $, where $g_j$ is the degeneracy of the $j-$th single-particle state, $j_{\mu}$ is the value of $j$ associated to the Fermi energy, and $\tilde{T} = k_B T / \varepsilon_0$ is the dimensionless effective temperature (with $\varepsilon_0$ being the energy of the lowest single-particle state, i.e. $j=0$). As it is expected, for small values of $z_x$, $j_{\mu} \thicksim N$ and $\tilde{T} \thicksim 0$. The temperature $\tilde{T}$ is a monotonic increasing function of $z_x$, while $j_{\mu} \thicksim N$ in a wide range of the $z_x$ parameter and it is a monotonic decreasing function of $z_x$ for $z_x$ sufficiently close to one. Then, the region in the parameter space for which $j_{\mu} \thicksim N$ could be a starting point to define a value or region of the parameter $z_x$ in which the fermionic character emerges (see Fig.~\ref{fig_DOS} (b) and (d)). Notice that this region becomes broader for larger $N$, the fermionic character arises for $z_x \thicksim 0.85$ for $N=10$ while for $N=100$ it arises for $z_x \thicksim 0.99$ (Fig.~\ref{fig_DOS} (b) and (d), respectively). 

We also performed a fitting of the DOS with the expression given by the Bose-Einstein (BE) distribution. We obtained that this distribution only holds for $z_x$ sufficiently close to one (larger values of $N$ require values of $z_x$ closet to one), case in which it gives the same values of $j_{\mu}$ and $\tilde{T}$ than those obtained with the FD distribution. If $z_x$ is not sufficiently close to one, the fitted BE distribution present significant deviations from the obtained DOS, in contrast, the FD fitted distribution shows a perfect agreement with the calculated DOS for any value of $z_x$. Therefore, we conclude that the one-dimensional system present a strong fermionic character for a wide range of the parameter $z_x$.
 
\begin{figure}[]
\includegraphics[width=0.85\columnwidth]{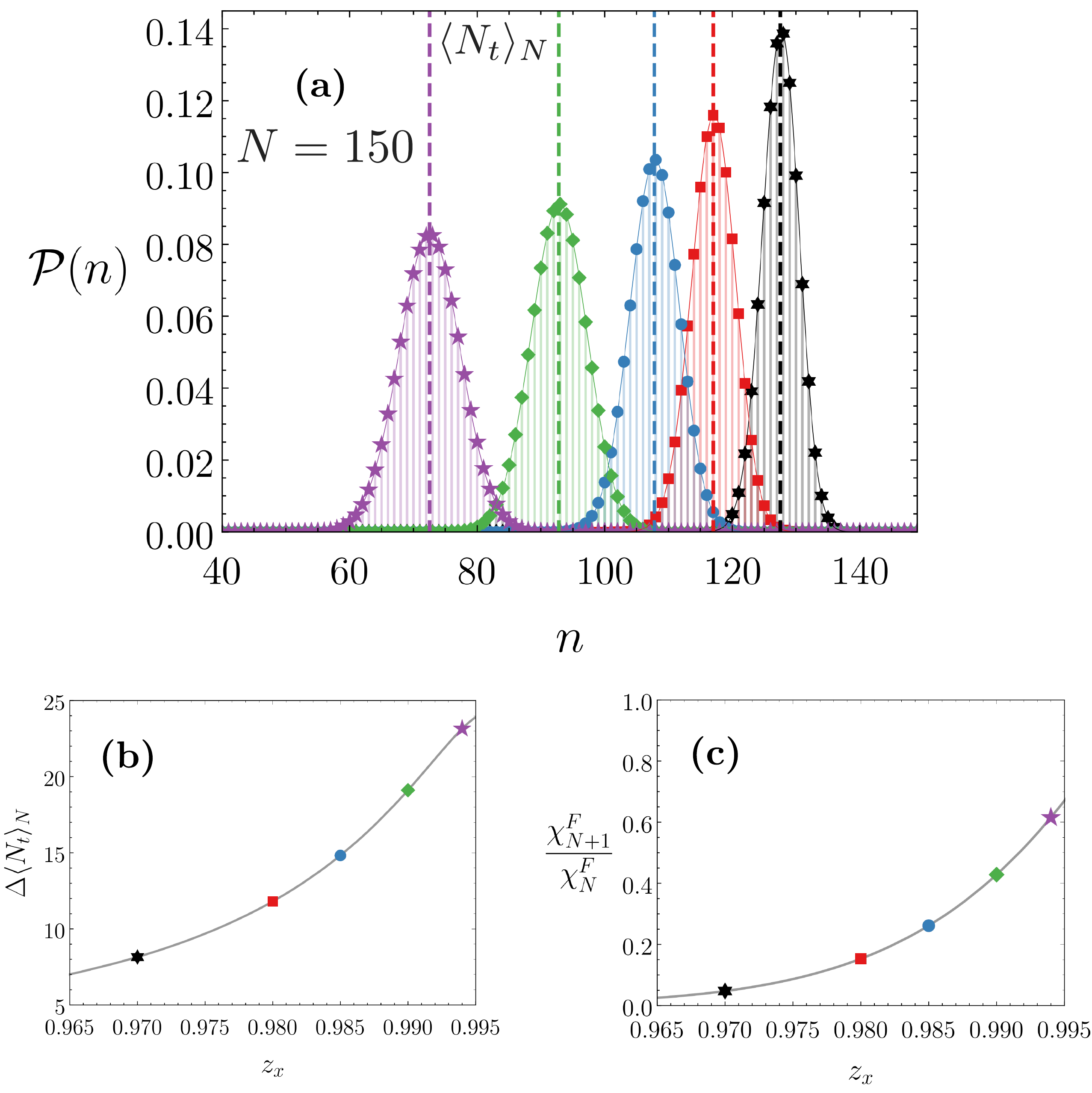}
\caption{Suppression of particle fluctuations due to Pauli blocking evidenced in the probability of finding $n$ fermionic pairs in the $t=N$ lowest energetic states, $\mathcal{P}(n)$ in panel (a). A system of $N=150$ confined particle pairs is presented for different values of $z_x$. The distributions are centered at $\mval{N_t}_N = \sum_{j=0}^{t-1} n_j$, depicted as the vertical dashed lines in the left panel. The variance of the distribution $\Delta \mval{N_t}_N$ is shown in panel (b) as a function of $z_x$. Black six-pointed stars, red squares, blue dots, green diamonds, and purple five-pointed stars indicate different values of $z_x$ chosen in order to cover completely the variation of the normalization ratio $\chi_{N+1}/\chi_N$ depicted in (c), i.e. to guarantee the inclusion of the regime in which the pairs behave as ideal bosons as well as the regime in which the fermionic character of the particles arises.}
\label{fig_prob}
\end{figure}

The presence of correlations due to the Pauli exclusion principle can also be evidenced in the probability of finding $n$ fermionic pairs in the $t$ lowest energetic states $\mathcal{P}(n)=\sum_{0\le j_1<j_2< \cdots < j_n\le t-1} \mval{\prod_{k=1}^n \hat a_{j_k}^\dagger \hat b_{j_k}^\dagger \hat a_{j_k} \hat b_{j_k}}_N = \binom{N}{n} \chi_{n}^{\tilde \Lambda_t} \chi_{N-n}^{\bar\Lambda_{S-t}}/\chi_N$, where $\chi_{n}^{\tilde \Lambda_t}$ and $\chi_{N-n}^{\bar\Lambda_{S-t}}$ are the normalization factors calculated with the first $t$ occupation numbers $\tilde  \Lambda_t = (\lambda_0,\lambda_1, \ldots,\lambda_{t-1})$ and the remaining ocuppations $\bar \Lambda_{\infty-t} = (\lambda_{t},\lambda_{t+1}, \ldots,\lambda_\infty)$ respectively, and $\chi_{n>t}^{\tilde  \Lambda_t } = 0$ prevents populations larger than $t$ (once more the $F$ superscript has been omitted for simplicity of notation). Due to Pauli blocking no more than $t$ fermionic pairs can populate this spectral region leading to a strong suppression of particle fluctuations \cite{Bouvrie2019}. This suppression is exposed by the decreasing variance of the probability distribution (see Fig.~\ref{fig_prob}) when decreasing $z_x$. In other words, the pairs of particles deviate from the bosonic behavior and the fermionic character of the particles arises. Notice that the probability $\mathcal{P}(n)$ is centered around the mean population of the $t$ lowest energetic states of the single-particle states, $\mval{N_t}_N = \sum_{j=0}^{t-1} n_j$. To sum up, for sufficiently small values of $z_x$ we obtained that $\mval{N_t}_N \thicksim N$ and $\Delta \mval{N_t}_N \thicksim 0$, in agreement with the step-like behavior of the DOS.     

\begin{figure}[]
\includegraphics[height=0.77\textheight]{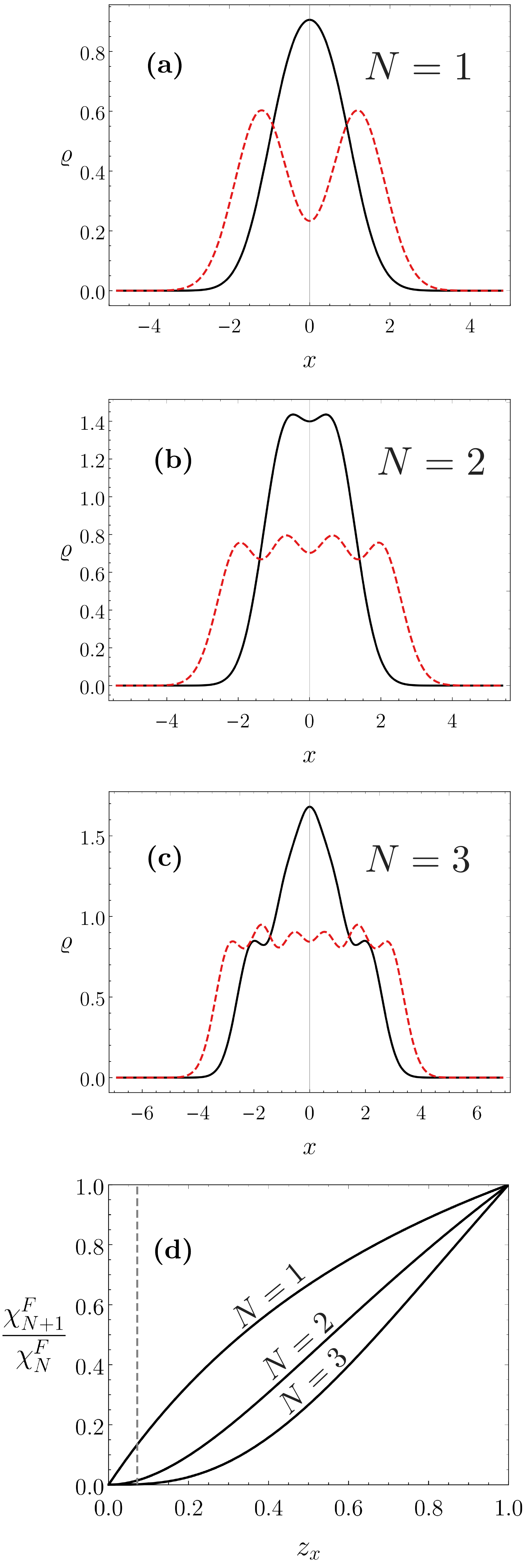}
\caption{Friedel-Wigner oscillations transition depicted by the total ground state density profile $\varrho(x)$ of $N$ confined fermionic pairs with Coulomb interactions for $N=1,\,2,\,3$ in panel (a), (b), and (c). The crossover is revealed by the doubling in the peaks of the total ground state density profile from $N$ (half of the total particle number) in the Friedel regime to $2N$ (total particle number) in the Wigner regime (black solid lines and dashed red lines respectively). The value of the $z_x$ parameter associated to the Coulomb interaction is shown as a gray dashed line in panel (d), where the normalization ratios are depicted. Notice that for this value of $z_x$ the cobosons present strong deviations from the ideal bosonic behavior.}
\label{fig_rho}
\end{figure}

Up to now we have presented a general characterization of a confined Wigner molecule constituted by an even number of fermions. In what follows, we would like to show that in the framework of the coboson ansatz it is possible to obtain similar results to those obtained by the authors of Refs. \cite{Xianlong2012, Wang2012, Soffing2012, Secchi2012, Kylanpaa2016, Cavaliere2014} studying the WMs within Quantum Montecarlo methods, exact diagonalization, density functional theory, and the Bethe ansatz formalism among others. Once more, we would like to stress the simplicity of our approach and its computational conveniences. We calculated the total ground state density profile $\varrho(x)= \mval{\hat \Psi^\dagger(x) \hat \Psi (x)}_N$ \cite{Wang2012} (with $\hat \Psi^\dagger(x)$ being the operator creating a particle in $x$) for $N$ pairs of confined fermions with interactions described by the Coulomb potential \cite{Kylanpaa2016}, see appendix~\ref{sec_appendix_2pWM}. Varying the interaction strength we are able to capture the Friedel-Wigner transition described in Refs. \cite{Xianlong2012, Kylanpaa2016}. As stated by Xianlong in Ref. \cite{Xianlong2012}, when the interaction between particles is not sufficiently strong the density is characterized by $2k_F$-Friedel oscillations with half of the number of particles peaks ($N$ in our present work). When the interaction strength increases the number of peaks in the density profile is doubled leading to $4k_F$-Wigner oscillations characterized by as many peaks as the number of particles ($2N$ in our case). The coboson ansatz is capable of capturing this crossover, as can be seen in figure~\ref{fig_rho} by the doubling of the peaks in the total ground state density profile when increasing the interaction strength between the two particles of different species that made up the composite boson. It is important to remark that for the value of $z_x$ associated to the Coulomb interaction, the composite bosons are far from the ideal bosonic regime (see Fig.~\ref{fig_rho} (d) and appendix~\ref{sec_appendix_2pWM}). Therefore, even when the fermionic character of the constituents of the composite boson becomes relevant, the ground state density profile calculated with the coboson ansatz shows close agreement with the results obtained when applying different methods. We conclude that our approach is able to faithfully reproduce the many particle physics of the system, not only when a strong degree of entanglement guarantees an ideal bosonic behavior, but also when the particles have strong correlations due to the Pauli exclusion principle. Since the coboson ansatz only considers the interactions given by fermionic exchanges, this also suggests that the Friedel-Wigner crossover arises mainly due to the exchange interactions.

\section{Bosonic behavior induced by symmetry in two dimensional Wigner molecules} 
\label{sec_res_2D}

The occupation numbers of the ground state of two interacting distinguishable particles confined in a two dimensional harmonic trap can be obtained in a closed form. This result can be extended to higher dimensions, as explained in Ref. \cite{Cuestas2017}. Based on these results it is possible to calculate the normalization ratio, the populations, the DOS, and the ground state density profile of a WM confined in two (or higher) dimensions. The two dimensional case is of particular interest due to the presence of Wigner crystals, while a similar study in three dimensions may constitute a strong foundation for a new approach to address ultracold interacting Fermi gases. Those issues will be addressed in a future work, here we focus on the proof of the bosonic induced behavior by symmetry. 

From the product form of the occupation numbers (see Eq.~\eqref{eq_occup}) it is easy to show that the power sums of order $m$  can be written as 

\eq
\label{eq_M_2D}
M^{2D}(m) = \frac{(1-z_x)^m}{\left(1- z_x^m\right)} \frac{(1-z_y)^m}{\left(1- z_y^m\right)} .
\en 

\noindent It is straightforward to see that $M^{2D}(1)=1$ and that $M^{2D}(m) \to 0$ if $m>1$ and $z_x \to 1$ or $z_y \to 1$ (the latter can be obtained by performing a very simple power series expansion). As mentioned in Sec.~\ref{sec_WM}, $z_x$ depends on the interaction strength and the parameters of the interaction potential, while $z_y$ depends on the anisotropy of the trap. The isotropic or symmetric trap is obtained for $z_y \to 1$, case in which $M^{2D}(m) \to 0$ for $m>1$. Now we rewrite Eq.~\eqref{eq_chis_sums} as

\begin{small}
\begin{eqnarray}
\label{eq_chis_sums_2}
\chi_{N}^F = \left(M(1)\right)^N + & & N! \sum_{k_1=0}^{N-1} \sum_{k_2=0}^{\floor{\frac{N}{2}}} \sum_{k_3=0}^{\floor{\frac{N}{3}}} \ldots \sum_{k_{N-1}=0}^{\floor{\frac{N}{N-1}}} \sum_{k_N=0}^1 \\
\nonumber
& &\delta_{k_1+2 k_2 + \ldots + N k_N , N} \frac{(-1)^{k_1 + k_2 + \ldots +k_N + N}}{k_1!k_2! \ldots k_N!}\\
\nonumber
& &\prod_{i=1}^{N} \left( \frac{M(i)}{i} \right)^{k_i} ,
\end{eqnarray}
\end{small}    

\noindent where $\floor{x}$ denotes the floor function, i.e. the largest integer less than or equal to $x$. The second term contains at least some factor $\left( \frac{M(m)}{m} \right)^{k_m}$ with non-zero $k_m$ and $m>1$. If $z_y \to 1$ we have $M^{2D}(m) \to 0$ for $m>1$, therefore, the second term becomes null and $\chi_{N}^{F\,2D} \to \left(M(1)^{2D}\right)^N \to 1$ for any number of particles. In other words, if $z_y \to 1$ the normalization ratio fulfills $\chi_{N+1}^{F\,2D}/\chi_N^{F\,2D} \to 1$ independent of the value of $z_x$. This means that for a symmetric or isotropic trap, no matter the range or strength of the interaction potential between the constituent parts of the molecule, the two distinguishable particles behave as an ideal boson. 

Another way to prove the ideal boson character induced by symmetry considers the upper and lower bounds derived by Chudzicki et al. in Ref. \cite{Chudzicki2010}, 

\eq
\label{eq_chi_bounds}
1 - N P \leqslant  \frac{\chi_{N+1}^{F}}{\chi_N^{F}}  \leqslant  1 - P \leqslant 1,
\en 

\noindent with $P$ being the purity of the state of Eq.~\eqref{eq_schmidt_form}, i.e. $P=M(2)$ and $1-P = \chi_2^{F} = S_L$, where $S_L$ denotes the linear entropy. In our case,

\eq
P^{2D}=\frac{(1-z_x)}{(1+z_x)} \frac{(1-z_y)}{(1+z_y)} ,
\nonumber
\en 

\noindent and then

\begin{small}
\eq
\label{eq_chi_bounds_2D}
1 - N \frac{(1-z_x)}{(1+z_x)} \frac{(1-z_y)}{(1+z_y)} \leqslant  \frac{\chi_{N+1}^{F\,2D}}{\chi_N^{F\,2D}}  \leqslant  1 -  \frac{(1-z_x)}{(1+z_x)} \frac{(1-z_y)}{(1+z_y)} , 
\en 
\end{small}

\noindent where we can explicitly see that $\chi_{N+1}^{F\,2D}/\chi_N^{F\,2D} \to 1$ for any value of $z_x$  whenever $z_y \to 1$.

Before concluding, we would like to stress that when considering composite bosons made up of two distinguishable bosons (instead of fermions) confined in a two dimensional harmonic trap, the normalization factor can be derived in a closed form and satisfies 

\eq
\label{eq_chi_bounds_2D_2}
\chi_N^{B\,2D} = \chi_N^{B\,1D} (z_x) \chi_N^{B\,1D} (z_y) , 
\en 

\noindent where $\chi_N^{B\,1D}(z)$ denotes the normalization factor found by Law in Ref. \cite{Law2005}. $\chi_{N+1}^{B\,1D}(z)/\chi_{N}^{B\,1D}(z)$ is always greater than one and it is equal to one only if $z=1$. Therefore, the two dimensional bosonic enhancement factor is greater than one for any pair of values of $z_x$ and $z_y$, and it goes to unity only if $z_{x}$ and $z_y$ are sufficiently close to one.  

\section{Summary} 
\label{sec_concl}

To summarize, we have studied Wigner molecules from a new and efficient approach. Our main focus was on determining if a Wigner molecule made up of $2N$ fermions behaves as $N$ bosons or $2N$ fermions. By applying the composite boson ansatz to $N$ pairs of harmonically confined fermions we were capable to address few particle Wigner molecules, as well as systems with a large number of particles. For few number of particles, we found that the wave function obtained in the framework of the coboson ansatz is capable to capture the Friedel-Wigner transition. For a large number of particles, we observed the suppression of particle fluctuations as a signature of strong fermionic correlations. 

We also presented more evidence supporting the hypothesis that the amount of entanglement between particles determines their bosonic composite degree. In this context, we discussed the physical meaning and the reason why a sufficiently large entanglement is needed in order to ensure that the particles behave as elementary bosons. We concluded that the composite behavior of confined particles depends on the availability of sufficient space in the state space (the effective number of Schmidt modes must be greater than the total number of composite particles) as well as on having enough available real space compared to its size. Finally, we also proved that a Wigner molecule confined in a two dimensional trap presents a bosonic behavior induced by symmetry regardless of the range or strength of the interaction potential between the constituent parts of the molecule. Therefore, as was pointed out by Law in Ref. \cite{Law2005}, 
the composite representation is not limited to strongly interacting particles but to particles with large enough degree of entanglement.

Finally, we would like to stress that our approach reproduces the many particle physics of a confined Wigner molecule not only when a strong degree of entanglement guarantees the ideal bosonic behavior of the composite particles, but also when the particles have strong correlations due to the Pauli exclusion principle. Then, the coboson ansatz becomes a powerful tool to address systems with strong fermionic correlations.

\begin{acknowledgements}

We thank J.I. Robledo for his careful reading of the manuscript. We acknowledge funding from grant PICT-BID 2017-2583 from Agencia Nacional de Promoción Científica y Tecnológica de Argentina (ANPCyT). E.C and A.P.M. acknowledge grant GRFT-2018 MINCYT-Córdoba, as well as the Argentinian agencies SeCyT-UNC and CONICET for their financial support. 
\end{acknowledgements}

\appendix

\section{Two-particle Wigner Molecules} 
\label{sec_appendix_2pWM}

In what follows we intend to summarize the main results of the two-particle Wigner molecule treated in Ref. \cite{Cuestas2017}. It provides the fundamental unity (a single coboson, Eq.~\eqref{eq_schmidt_form}) from which the $2N-$ particle state is constructed within the coboson ansatz (Eq.~\eqref{eq_ket_N}). In this work we are considering two distinguishable particles, therefore there are some subtle differences between the results presented here compared to those of Ref. \cite{Cuestas2017}. In the mentioned work the authors obtain symmetric or antisymmetric states, here we are looking for states in which the two distinguishable particles are accessible independently \cite{Eckert_2002}. In other words, we need to found the ground state of the system in which the two distinguishable particles can be locally addressed. When considering such states the occupation numbers do not present the double degenerancy mentioned in Ref. \cite{Cuestas2017}, and the normalization constant is simpler. The entanglement entropies can be derived following the same procedures and lead to the same functional expressions up to a $1/2$ factor not appearing in the present case.

Let us start by considering two-interacting particles in a two-dimensional anisotropic trap,

\begin{small}
\begin{eqnarray*}
\label{H_QD_2D}
{\cal H} = &-&\frac{\hbar^2}{2 m}\left( \nabla_1^2+\nabla_2^2\right) + \frac{ m \omega^2}{2}\left( (x_1^2+x_2^2)+\varepsilon^2(y_1^2+y_2^2) \right) + g {\cal V} (r) ,
\end{eqnarray*}
\end{small}

\noindent with $\varepsilon > 1$ being the anisotropy parameter, $m$ the mass of the particles, ${\cal V}(r)$ the interaction potential, $r$ the distance between particles, and $g$ the interaction strength. By introducing the center of mass $\vec{R} = \frac{1}{2} (\vec{r}_1 + \vec{r}_2)$ and relative coordinates $\vec{r} = \vec{r}_2-\vec{r}_1$ the Hamiltonian decouples as  ${\cal H} = {\cal H}^R + {\cal H}^r$, where

\begin{eqnarray*}
\label{H_cal_2D_mc}
& &{\cal H}^R = -\frac{\hbar^2}{2 m_R} \nabla_R^2 + \frac{ m_R \omega^2}{2} R^2 , \\
\label{H_cal_2D_r}
& &{\cal H}^r = -\frac{\hbar^2}{2 m_r} \nabla_r^2 + \frac{ m_r \omega^2}{2} r^2 + g {\cal V} (r) , 
\end{eqnarray*}

\noindent with $m_r = m/2 $ and $m_R = 2 m$. The total wave function is the product of the center of mass wave function and the relative wave function $\Psi(r,R) = \psi^R(R) \psi^r(r)$, where the center of mass states are the harmonic oscillator ones. In order to obtain the relative wave functions the authors of Ref. \cite{Cuestas2017} consider the strong interacting regime and present a method based on the harmonic approximation of the potential ${\cal V} (r)$. If the potential is repulsive, decreases monotonously, $ {\cal V} ( r ) \to 0 $ for $r\to\infty$, and $\varepsilon > 1$, then, the minima of the potential locate on the $x-$axis $\vec{r}_{min} =\left(\pm x_0, 0 \right)$ with $x_0>0$ given by

\begin{equation}
\label{min_V_eff}
\frac{m_r \omega^2}{2g}=-\left.\left( \frac{1}{r} \frac{\partial {\cal V}}{\partial 
r}\right)\right\vert_{x_0} . 
\end{equation}

This procedure leads to a Hamiltonian of uncoupled oscillators,

\begin{equation*}
\label{H_Ha}
{\cal H}^r_{HA} = -\frac{\hbar^2}{2 m_r} \nabla_r^2 + \frac{m_r \omega^2}{2} \left\lbrace \mu^2 \left( x- x_0  \right)^2 + \left(\varepsilon^2 -1 \right) y^2 \right\rbrace , 
\end{equation*}

\noindent where 

\begin{equation}
\label{wx_2}
\mu^2 = \frac{\left. \frac{\partial^2 {\cal V}}{\partial 
r^2}\right\vert_{x_0} }{\left. -\frac{1}{r} \frac{\partial {\cal V}}{\partial 
r}\right\vert_{x_0}} + 1 ,
\end{equation}

\noindent notice that the dependence on the interaction strength $g$ and  the parameters of the potential is implicit in $x_0$. It is also important to keep in mind that these results are valid in the strong interacting regime, in which the ratio between the interaction strength and the confining energy is sufficiently large.    

Now we are able to obtain the ground state wave function. As mentioned earlier, in Ref. \cite{Cuestas2017} the authors obtain symmetric or antisymmetric states, but here we are interested in the ground state of the system in which the two distinguishable particles can be locally addressed. Such state is also a product of Gaussians states and the same procedure given in the supporting information of Ref. \cite{Cuestas2017} can be applied in order to obtain the occupations, 

\eq
\label{eq_occup_rep}
\nonumber
\lambda_{j_x, j_y}^{2D} = \lambda_{j_x} \lambda_{j_y} = (1-z_x) z_x^{j_x} (1-z_y) z_y^{j_y} ,
\en 

where

\eq
\label{eq_zx}
z_x = \left( \frac{1-\mu}{1+\mu} \right)^2 ,
\en 

and

\eq
\label{eq_zy}
z_y = \left( \frac{\left( \varepsilon^2 -1 \right) ^\frac{1}{4} - \sqrt{\varepsilon}}{\left( \varepsilon^2 -1 \right) ^\frac{1}{4} + \sqrt{\varepsilon}}\right)^2 .
\en 

As mentioned in the main text $j_{x,y}=0,1,2,\ldots, \infty$ gives all the spectrum of the one-particle reduced density matrix and $0<z_{x,y}<1$. Notice that $z_x$ depends on the ratio between the interaction strength and the confining energy as well as on the parameters of the interaction potential, while $z_y$ depends on the anisotropy of the trap. The one dimensional case is re-obtained by taking $z_y \to 0$ and the isotropic trap is obtained for $z_y \to 1$. 

Different measures of entanglement are given in terms of the distribution of the occupations. In Ref. \cite{Cuestas2017} the family of R\'enyi entropies ($S^\alpha$), the linear ($S_L$), and von Neumann ($S_{vN}$) entropies are calculated, as well as the min- ($S^\infty$) and Hartley or max- ($S^0$) entropy as limiting cases within the family of R\'enyi entropies. Due to the product form of the occupations, the R\'enyi and von Neuman entropies can be expressed as a sum of two terms,

\begin{eqnarray*}
\label{Renyi_xy}
S^{\alpha}_{2D} =& & \, S^{\alpha}_{x}(z_x) + S^{\alpha}_{y}(z_y) \\
\nonumber
= & & \, \frac{1}{1-\alpha} \left( \log_2 \left(  
\frac{(1-z_x)^ \alpha}{(1-z_x^\alpha)}
\right) +  \log_2 \left(  
\frac{(1-z_y)^ \alpha}{(1-z_y^\alpha)}
\right) \right) ,
\end{eqnarray*}

\begin{eqnarray*}
\label{S_VN}
S_{vN}^{2D} =  \, & & S_{x}^{1}\left( z_x \right) + S_{y}^{1}(z_y) \\
\nonumber
= & & \, - \frac{\log_2\left( \left(1- z_x \right)^{(1- z_x )} z_x^{z_x} \right)}{ \left(1-z_x\right)}  - \frac{\log_2\left( \left(1- z_y \right)^{(1- z_y )} z_y^{z_y} \right)}{ \left(1-z_y\right)},
\end{eqnarray*}

\begin{eqnarray*}
\label{S_min}
S^{\infty}_{2D} =  & & \lim_{\alpha\to\infty} \left( S^{\alpha}_{x}\left( z_x \right) + S^{\alpha}_{y}(z_y) \right) = S^{\infty}_{x}\left( z_x \right)+S^{\infty}_{y}(z_y), 
\end{eqnarray*}

\noindent and,

\begin{eqnarray*}
\label{S_max}
S^{0}_{2D} =  & & \lim_{\alpha\to 0} \left( S^{\alpha}_{x}\left( z_x \right) + S^{\alpha}_{y}(z_y) \right) = S^{0}_{x}\left( z_x \right)+S^{0}_{y}(z_y) . 
\end{eqnarray*}

Finally, the linear entropy is

\eq
\nonumber
S_L^{2D} = 1 - \frac{(1-z_x)}{(1+z_x)} \frac{(1-z_y)}{(1+z_y)} ,
\nonumber
\en 

\noindent which reaches its maximum value in the isotropic case $z_y \to 1$ or, in the one dimensional case ($z_y \to 0$) it goes to one if $z_x \to 1$.

The family of R\'enyi entropies and the von Neumann entropy present a logarithmic divergence in the isotropic case, $z_y \to 1$. They also diverge in the one dimensional case, $z_y \to 0$, when $z_x \to 1$.  

It is important to notice that $z_x \thicksim 1$ if $\mu \gg 1$. In Ref. \cite{Cuestas2017} the authors show that in the strong interaction strength regime, this condition is reached for short range potentials. Based on the discussion presented in Sec.~\ref{sec_WM}, this means that the two-particle system would present a strong bosonic character if the interaction between the particles has short range, this is in agreement to what was stated by Combescot for ultracold interacting Fermi gases \cite{CombescotShiau2015,CombescotShiauChang2016}. For long range potentials, as the Coulomb potential (see Fig.~\ref{fig_rho} of Sec.~\ref{sec_res_1D}), $z_x$ is not close to one and the fermionic character of the particles becomes relevant in the many particle physics as it is discussed in the main text, even so, the composite boson ansatz is capable of recovering features as the Friedel-Wigner crossover. 

Here we focus on long range potentials because the most realistic and widely used interparticle interactions for studying Wigner molecules confined in quantum dots are the long range Coulomb or softened Coulomb potentials (see \cite{Kylanpaa2016, Wang2012, Cavaliere2014,pecker_2013}, and references therein). Then, we particularize the results with the inverse power interaction, ${\cal V} (r) = 1/r^{\gamma}$, for which the Coulomb potential used in Sec.~\ref{sec_res_1D} is a particular case. For this potential, Eq.~\eqref{min_V_eff} gives $x_0 = \pm \sqrt[\gamma+2]{2 \gamma g/ m_r \omega^2 }$, where we can see that the minima is an increasing function of the interaction strength $g$. Even more, from Eq.~\eqref{wx_2} $\mu^2 = \gamma +2$ is obtained, therefore, for a finite value of $\gamma$ the value of the parameter $z_x$ remains below one, as it is shown in Fig.~\ref{fig_rho} (d).

Before ending this appendix we would like to write down the explicit expressions obtained for the one dimensional case by taking $z_y \to 0$ in the previous results. The one-dimensional occupations are

\eq
\label{eq_occup_1D}
\nonumber
\lambda_{j_x}^{1D} = \lambda_{j_x} = (1-z_x) z_x^{j_x},
\en 

\noindent with $z_x$ as in Eq.~\eqref{eq_zx}. The family of R\'enyi entropies, the linear, von Neumann, and the min- and Hartley or max- entropy are

\begin{eqnarray*}
\label{Renyi_xy_1D}
S^{\alpha}_{1D} =& & \, S^{\alpha}_{x}(z_x) \\
\nonumber
= & & \, \frac{1}{1-\alpha}  \log_2 \left(  
\frac{(1-z_x)^ \alpha}{(1-z_x^\alpha)}
\right) ,
\end{eqnarray*}

\begin{eqnarray*}
\label{S_VN_1D}
S_{vN}^{1D} =  \, & & S_{x}^{1}\left( z_x \right) \\
\nonumber
= & & \, - \frac{\log_2\left( \left(1- z_x \right)^{(1- z_x )} z_x^{z_x} \right)}{ \left(1-z_x\right)} ,
\end{eqnarray*}

\begin{eqnarray*}
\label{S_min_1D}
S^{\infty}_{1D} =  & & \lim_{\alpha\to\infty} S^{\alpha}_{x}\left( z_x \right)  = S^{\infty}_{x}\left( z_x \right) , 
\end{eqnarray*}

\begin{eqnarray*}
\label{S_max_1D}
S^{0}_{1D} =  & & \lim_{\alpha\to 0}  S^{\alpha}_{x}\left( z_x \right)  = S^{0}_{x}\left( z_x \right) , 
\end{eqnarray*}
 
\noindent and,

\eq
\nonumber
S_L^{1D}= 1 - \frac{(1-z_x)}{(1+z_x)} .
\nonumber
\en 

We conclude this appendix with a remark concerning the $z_x$ and $z_y$ parameters. As we mentioned before, $z_x$ depends on the ratio between the interaction strength and the confining energy as well as on the parameters of the interaction potential (see Eq.~\eqref{wx_2} and \eqref{eq_zx} ), and $z_y$ depends on the anisotropy of the trap (see Eq.~\eqref{eq_zy}). Therefore, given that in the experiments the form and range of the interaction is fixed, a possible way to control the $z_x$ parameter might be by controlling the ratio between the interaction strength and the confining energy, i.e. by changing for instance the effective width of the trap. On the other hand, the $z_x$ parameter (which already appears in the one-dimensional case) remains unchanged when considering two or higher dimensions provided that the interaction potential and the trap remain unchanged. In this sense, the influence of the interaction potential is already and fully present in dimension one, and the $d$-dimensionality contributes with $d-1$ parameters of anisotropy $\varepsilon_i$ with $i = 1, \ldots, d-1$, or equivalently, $d-1$ parameters $z^{i}_y(\varepsilon_i)$ where $i = 1, \ldots, d-1$. For the sake of completeness, we give the explicit expressions for the $d-$dimensional case (see supplementary material of Ref. \cite{Cuestas2017}). The occupations:

\eq
\label{eq_occup_DD}
\nonumber
\lambda_{j_x, \lbrace j_{y_i} \rbrace}^{dD} = \lambda_{j_x}  \prod_{i=1}^{d-1} \lambda_{j_{y_i}} = (1-z_x) z_x^{j_x} \prod_{i=1}^{d-1} (1-z_{y_i}) z_{y_i}^{j_{y_i}},
\en 

\noindent with $z_x$ as in Eq.~\eqref{eq_zx}, and

\eq
\label{eq_zyi}
z_{y_i} = \left( \frac{\left( \varepsilon_i^2 -1 \right) ^\frac{1}{4} - \sqrt{\varepsilon_i}}{\left( \varepsilon_i^2 -1 \right) ^\frac{1}{4} + \sqrt{\varepsilon_i}}\right)^2 .
\en 

\noindent for $i=1,\ldots,d-1$. The family of R\'enyi entropies, the linear, von Neumann, and the min- and Hartley or max- entropy:

\begin{eqnarray*}
\label{Renyi_xy_dD}
S^{\alpha}_{dD} =& & \, S^{\alpha}_{x}(z_x) + \sum_{i=1}^{d-1} S^{\alpha}_{y}(z_{y_i}) \\
\nonumber
= & & \, \frac{1}{1-\alpha} \left( \log_2 \left(  
\frac{(1-z_x)^ \alpha}{(1-z_x^\alpha)}
\right) +  \sum_{i=1}^{d-1} \log_2 \left(  
\frac{(1-z_{y_i})^ \alpha}{(1-z_{y_i}^\alpha)}
\right) \right) ,
\end{eqnarray*}

\begin{eqnarray*}
\label{S_VN_dD}
S_{vN}^{dD} = \, & & S_{x}^{1} \left( z_x \right) + \sum_{i=1}^{d-1} S_{y}^{1}(z_{y_i}) \\
\nonumber
= & & \, - \frac{\log_2\left( \left(1- z_x \right)^{(1- z_x )} z_x^{z_x} \right)}{ \left(1-z_x\right)}  - \sum_{i=1}^{d-1} \frac{\log_2\left( \left(1- z_{y_i} \right)^{(1- z_{y_i} )} z_{y_i}^{z_{y_i}} \right)}{ \left(1-z_{y_i}\right)},
\end{eqnarray*}

\begin{eqnarray*}
\label{S_min_dD}
S^{\infty}_{dD} =  & & \lim_{\alpha\to\infty} \left( S^{\alpha}_{x}\left( z_x \right) + \sum_{i=1}^{d-1} S^{\alpha}_{y}(z_{y_i}) \right) = S^{\infty}_{x}\left( z_x \right)+ \sum_{i=1}^{d-1} S^{\infty}_{y}(z_{y_i}), 
\end{eqnarray*}

\begin{eqnarray*}
\label{S_max_dD}
S^{0}_{dD} =  & & \lim_{\alpha\to 0} \left( S^{\alpha}_{x}\left( z_x \right) + \sum_{i=1}^{d-1} S^{\alpha}_{y}(z_{y_i}) \right) = S^{0}_{x}\left( z_x \right)+ \sum_{i=1}^{d-1} S^{0}_{y}(z_{y_i}) , 
\end{eqnarray*}
 
\noindent and,

\eq
\nonumber
S_L^{dD}= 1 - \frac{(1-z_x)}{(1+z_x)} \prod_{i=1}^{d-1} \frac{(1-z_{y_i})}{(1+z_{y_i})} .
\nonumber
\en 


\bibliography{cob_WM}
\bibliographystyle{h-physrev5}


\end{document}